# Dynamics of large-scale solar-wind streams obtained by the double superposed epoch analysis. 4. Helium abundance

Yu. I. Yermolaev[1,2], I. G. Lodkina[1], M. Yu. Yermolaev[1], M. O. Riazantseva[1], L.S. Rakhmanova[1]
N. L. Borodkova[1], Yu. S. Shugay[3], V. A. Slemzin[4], I. S. Veselovsky[1,3], D. G. Rodkin[4]

[1] Space Research Institute (IKI), Russian Academy of Sciences, Moscow, Russia

[2] Moscow Institute of Physics and Technology, Dolgoprudny, Russia

[3] Skobeltsyn Institute of Nuclear Physics, Lomonosov Moscow State University, Moscow, Russia

[4] P.N. Lebedev Physical Institute, Russian Academy of Sciences, Moscow, Russia

## Abstract

This work is a continuation of our previous articles (Yermolaev *et al.* in *J. Geophys.Res*. **120**, 7094, 2015; Yermolaev *et al.* in *Solar Phys.* **292**, 193, 2017; Yermolaev *et al.* in *Solar Phys.* **293**, 91, 2018), which describe average temporal profiles of interplanetary plasma and field parameters in large-scale solar-wind (SW) streams: corotating interaction regions (CIRs), interplanetary coronal mass ejections (ICMEs including both magnetic clouds (MCs) and ejecta), and sheaths as well as interplanetary shocks (ISs). In this work we analyze the average profile of helium abundance $N\alpha/Np$ for interval of 1976-2016. Our results confirm the main earlier obtained results: $N\alpha/Np$ is higher in quasi-stationary fast streams than in the slow ones; it slowly changes in compression regions CIRs and Sheaths from values in undisturbed solar wind to values in the corresponding type of a piston (HSS or ICME); in Ejecta it is closed to abundance in undisturbed streams and it is maximal in MCs. We obtained for the first time that $N\alpha/Np$ correlates with plasma β-parameter in compression regions CIRs and Sheaths and anti-correlates in ICMEs. The $N\alpha/Np$ *vs.* *β* dependence is stronger in MCs than in Ejecta and may be used as an indicator of conditions in the CME formation place on the Sun.

## 1. Introduction

An experimental study of the minor ion components of the solar wind is of great interest for two main reasons. First, their mass and ionization composition does not change after the solar wind leaves the corona in the interplanetary space, and therefore their measurements at 1 AU allow us to obtain information about the conditions for the formation of solar wind structures on the Sun (*e.g.*, Zurbuchen *et al.*, 1999, 2016; Kasper *et al.*, 2012; Rod'kin *et al.*, 2016; Zhao *et al.*, 2017; Fu *et al.*, 2018 and references therein). Secondly, due to their great diversity in mass and charge and their small concentration, they can be considered as test particles in the development of various physical processes in the phenomena of the solar wind and the magnetosphere (*e.g.*, Borovsky, 2008; Šafránková *et al.*, 2013; Broll *et al.*, 2017 and references therein). The doubly ionized helium ions $^4He^{++}$ (α-particles), the most abundant of minor ions, are most often used to analyze the solar wind since the 1970s (*e.g.*, reviews by Neugebauer, 1981, Veselovskii, 1984; Ermolaev, 1994).

The solar wind has two characteristic scales at 1 AU: (1) in the region of large scales, the solar wind retains information about solar atmosphere, and (2) in the region of small scales, the solar wind properties depends on local turbulence of the medium (*e.g.*, Zelenyi and Milovanov, 2004 and references therein). Borovsky (2008) concluded on the basis of experimental data that the solar wind consists of the magnetic flux tubes with the mean size ~ 4.4 $10^4$ km (the mean duration ~$10^3$ s) formed on the Sun and the helium abundance, $N\alpha/Np$, jumps at the walls of the tubes. Recently, Šafránková *et al.* (2013) described abrupt jumps of *Nα/Np* at the scale of 3 s (time resolution of plasma instrument) and suggested that the jumps are generated by the local kinetic processes in the turbulent solar wind. These measurements are in good agreement with our estimation (~$10^2$ s) of scale boundary between solar- and local-induced phenomena in the solar wind (Yermolaev, 2014). In this article, we use 1-h data of OMNI2 dataset and, therefore, we study the phenomena being born on the Sun.

As has been shown by numerous experiments (*e.g.* Ermolaev, 1994), the average helium abundance, *Nα/Np*, increases with increasing solar wind bulk proton speed, *Vp*. This fact was explained by the predominant acceleration of helium ions by Alfven waves, $V_A$, in the solar corona, since the magnitude of the Alfvén velocity is higher in fast SW streams than in slow ones. Wang [2008, 2016] found that the near-Earth helium abundance is an increasing function of the magnetic field strength, B, and Alfvén speed, $V_A$, in the outer corona. One of another possible physical mechanisms of heavy ion escape from the Sun to the interplanetary space is their Coulomb friction with the main, proton, component of the plasma when the minimal proton flux, exceeds a threshold (Geiss *et al.*, 1970). Gosling *et al.* (1981) suggested an empirical model based on the measurements on Vela and IMP satellite series, and this model predicts that the helium abundance anti-correlates with the density, *Np*, in slow SW streams from the coronal streamers. The average dependences of *Nα/Np* on the flux and density of the solar wind show an average diminishing trend, but the results of some experiments on selected intervals of observations and in narrow ranges of flux and density do not correspond to this average dependence (see, for example, Fig. 3 in the review by Ermolaev (1994)). The difference between the dependences has been explained by the experimental facts that the dependences are different in different large-scale SW phenomena. These facts allowed one to use the minor ion measurements for identification of SW events and their possible sources on the Sun. For instance, the relative helium abundance *Nα/Np* was suggested for selection of different solar wind large-scale phenomena in articles by Yermolaev (1990, 1991). Helium abundance in the slow streams from the coronal streamer belts (including the Heliospheric Current Sheet, HCS) is lower than in the fast High-Speed Streams (HSS) from the coronal holes. The maximal values of *Nα/Np* is observed in the Interplanetary Coronal Mass Ejections (ICMEs including Magnetic Clouds (MCs) with high rotating magnetic field and Ejecta without regular structure of field) and helium abundance is higher in MCs than in Ejecta. The *Nα/Np* values in both types compression regions (Corotating Interaction Regions (CIRs) before HSSs and Sheaths before fast ICMEs) is supposed to be intermediate between undisturbed slow streams before compression regions and corresponding fast piston (HSS or ICME).

Recently we calculated the average temporal profiles of several interplanetary and magnetospheric parameters for eight usual sequences of SW phenomena: (1) SW/CIR/SW, (2) SW/IS/CIR/SW, (3) SW/Ejecta/SW, (4) SW/Sheath/Ejecta/SW, (5) SW/IS/Sheath/Ejecta/SW, (6) SW/MC/SW, (7) SW/Sheath/MC/SW, and (8) SW/IS/Sheath/MC/SW (where SW is undisturbed solar wind, IS means interplanetary shock) for interval 1976-2000 (Yermolaev *et al.,* 2015). Figure 1 shows that the helium abundance measurements are very rare during period 1976-2000, and their results could not be used for selection of SW types. Nevertheless we have made calculations of the average temporal profiles of helium abundance for 8 sequences of SW phenomena for 1976-2000 interval. Analysis of these results showed that a large part of them has insufficient statistical significance because of high variability of *Nα/Np* and low number of its measurements and its profiles have not been included in our first publication (Yermolaev et al., 2015). Now we have prepared an extended catalog of large-scale SW phenomena up to 201**7**, which contains two times more events. In this article we analyze two versions of extended data set. Initially we submitted results for interval of 1976-2016 received from OMNI2 dataset website in the beginning of 2018 (red circles in Figure 1). On November 06, 2018 ACE Alpha/Proton density ratio, $N\alpha/Np$, was removed from OMNI2 (An explanation on the site is following: "because there was a very poor correlation between the Wind/SWE/NLF Alpha/Proton density ratios and ACE/SWEPAM", see details at https://omniweb.sci.gsfc.nasa.gov/html/ow_data.html#norm_pla). We repeated calculations for new OMNI2 dataset without ACE data for interval 1976-2017 (green rhombs in figure 1). Comparison of results obtained with two datasets allows one to study the contribution of ACE $N\alpha/Np$ data in statistic investigation of large-scale solar-wind phenomena.

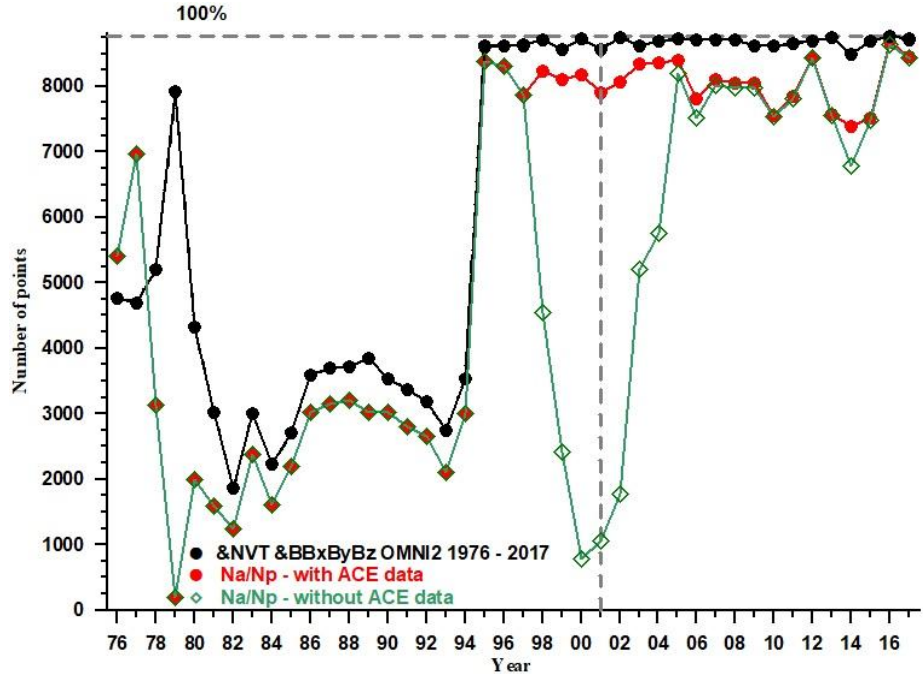

**Fig. 1.** *Year numbers of 1-hour data in the OMNI2 archive for interval 1976-2017: (a) plasma (density N, velocity V and temperature T) and magnetic field (IMF magnitude and components) data (black points), (b) relative helium abundance Nα/Np with ACE data (red points), and (c) relative helium abundance Nα/Np without ACE data (green rhombs). The data to the left of the dashed line were used in previous publications (Yermolaev et al., 2015; 2017; 2018).*

The main aim of this work is the study of $N\alpha/Np$ temporal profiles for the total period of measurements 1976-2017. The structure of the article is as follows. In section 2, the used data and methods are described. The obtained results are presented in section 3. Section 4 discusses and summarizes the results.

**2. Data and Methods**

In present paper, we use the same database and methods as in our previous works (Yermolaev *et al.*, 2015; 2017; 2018): (1) the one hour (1-h) interplanetary plasma and magnetic field data of the OMNI2 database (http://omniweb.gsfc.nasa.gov (King and Papitashvili, 2004)), (2) our extended catalog of large-scale solar-wind phenomena during 1976-2016 (ftp://ftp.iki.rssi.ru/pub/omni/ (Yermolaev *et al.*, 2009)) and (3) the double superposed epoch analysis method (Yermolaev *et al.*, 2010). This method involves re-scaling (proportional increasing/decreasing time between points) the duration of the interval for all SW types in such a manner that, respectively, beginnings and ends of all intervals of a selected type coincide. Similar methods of profile analysis were used in the papers by Yokoyama and Kamide (1997); Lepping *et al.* (2003, 2017).

As have been indicated in the Introduction, in this work we use an extended set of SW events. In contrast with our previous works (Yermolaev *et al.*, 2015; 2017; 2018 ), where we analyzed the 25-year set for period of 1976-2000, here we study the 41-year set with ACE $N\alpha/Np$ data for period of 1976-2016 and the 42-year set without ACE $N\alpha/Np$ data for period of 1976-2017 (see Figure 1 for comparison of availability of OMNI2 data for these periods). As a result, the number of events in the new 41-year and 42-year data sets turned out to be > 2 times more than in the old 25-year set. For 41-year set with ACE $N\alpha/Np$ data, we excluded several events which have not measurements of helium abundance during the whole intervals of SW events, and our analysis of this set includes 944 events for CIR, 706 for Sheath, 1191 for Ejecta, and 147 for MC. To maintain high statistics in the 42-year dataset after deleting ACE helium measurements, we (1) included 2017 data in the analysis, and (2) included events for which there was no complete continues series of helium measurements. As results, the 42-year dataset after deleting ACE helium measurements includes 1058 events for CIR, 849 for Sheath, 1407 for Ejecta, and 183 for MC. It should be noted that, although the number of events in the 42-year set is slightly higher than in the 41-year set, the number of 1-hour points in both sets is approximately equal, since the 42-year set contains events that have incomplete data sets on helium.

The increase in numbers of events did not change the standard deviation of helium abundance $N\alpha/Np$, σ, in the selected SW types but significantly decreased the statistic error, SE = σ/√N (N is number points in data set). Only for MCs and Sheaths before MCs SE = 0.007 - 0.01 and for all other solar wind types SE < 0.003. Small statistic errors allow us to make a reliable comparison of the average $N\alpha/Np$ values in different SW types. Because of large spread of data in Figure **3** the magnitudes of the correlation coefficients, |r|, change in a wide interval from 0.03 up to 0.59. Nevertheless the standard procedure (Bendat and Piersol, 1971) shows that the statistical significances for linear approximations with |r| > 0.5 are high and the probabilities for these cases P = 0.96 -- 0.99.

## 3. Results

In this section, we successively present the results of processing two extended data sets: first, the 41-year set with helium measurements on an ACE satellite, and then the 42-year set without ACE data.

Figure 2 presents the temporal profiles of helium abundance $N\alpha/Np$ (red line) and proton $\beta$-parameter (ratio of proton thermal to magnetic pressures - black line) obtained by the double superposed epoch analysis method (Yermolaev *et al.*, 2010) for the 41-year set with helium measurements on an ACE satellite. We show the $\beta$-parameter in this figure for two reasons. First, the dimensionless $\beta$-parameter is one of the main criteria in identifying large-scale SW types, and its presence in the figure makes it possible to clearly show the intervals of various SW types (*e.g.* Zurbuchen and Richardson, 2006; Yermolaev et al, 2009; 2015). Secondly, the temporal profile of the $\beta$-parameter was found to be related to the profile of the helium abundance $N\alpha/Np$, and this dynamics depends on the SW type. To show the latter, we performed the following procedure. On each panel and for each SW type in Figure 2 we compare the 1st point of abundance $N\alpha/Np$ with the 1st point of $\beta$-parameter, the 2nd point of abundance with the 2nd point of β-parameter, and so on. As a result of this comparison, the *Nα/Np* vs. *β*-parameter dependences are obtained, and these dependences are shown in Figure 3.

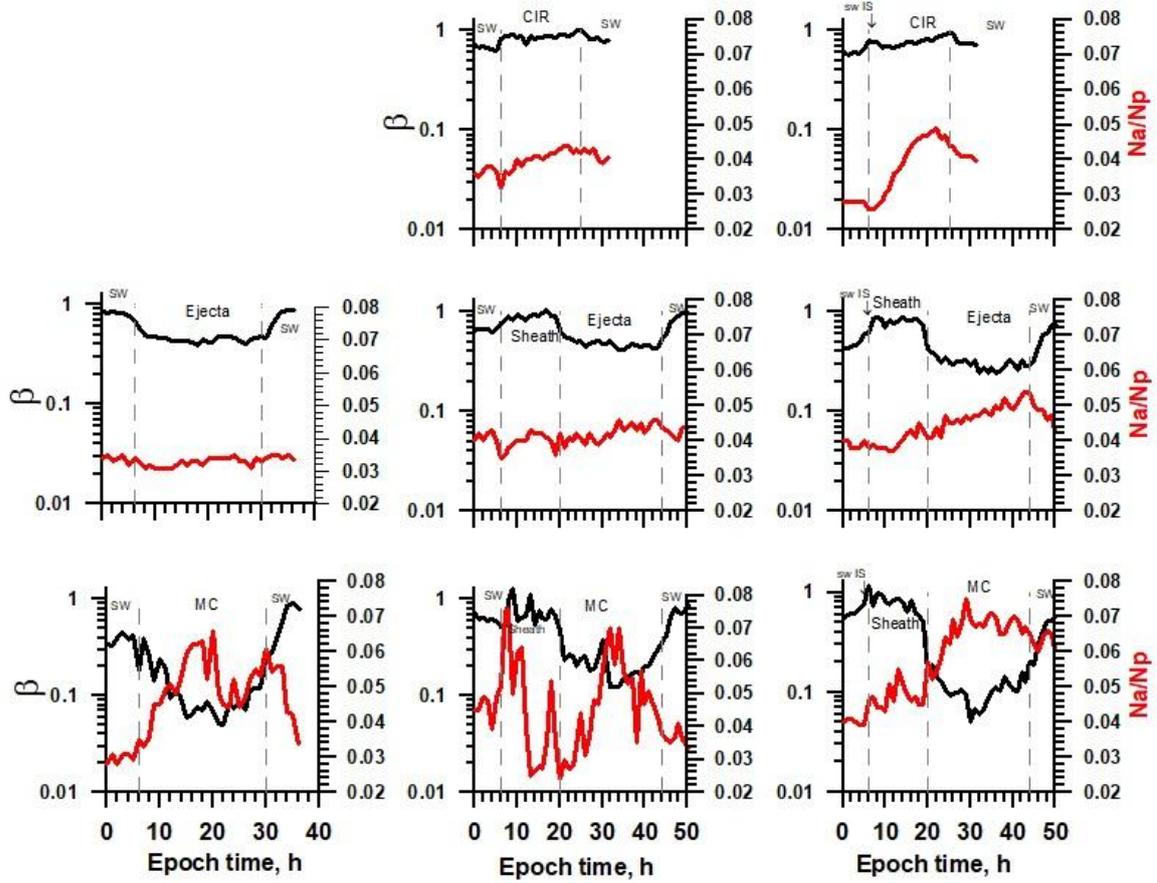

**Fig. 2.** *Time variations of helium abundance Na/Np (red line, right vertical axis) and proton β-parameter (black line, left vertical axis) obtained by the double method of superposed epoch analysis in eight sequences of SW events for the 41-year set with helium measurements on an ACE satellite: (1) CIR, (2) IS/CIR, (3) Ejecta, (4) Sheath/Ejecta, (5) IS/Sheath/Ejecta, (6) MC, (7) Sheath/MC, and (8) IS/Sheath/MC. Dashed vertical lines show boundaries of SW types which are indicated in the upper parts of each panel.*

Data in Figure 3 have a large spread. Nevertheless, the data in several panels have sufficiently large magnitude of correlation coefficients |r| > 0.5 and the high statistical significances for linear approximations with the probabilities P = 0.96 -- 0.99 (see Section 2). Data in these panels indicate clear trends: in compression regions CIR and Sheath the *Nα/Np* vs. *β*-parameter trends are positive and in ICME the trends are negative. The trend for MCs is stronger than for Ejecta. These trends for cases with |r| > 0.5 are observed in both cases: with and without shocks. Approximations for MCs with and without ISs are *Nα/Np* = 0.075 - 0.085 *β* and *Nα/Np* = 0.066-0.087 *β*, respectively, i.e. the slopes for both approximations are close, and the abundance at the minimum value of β is ~0.01 higher for the case with ISs.

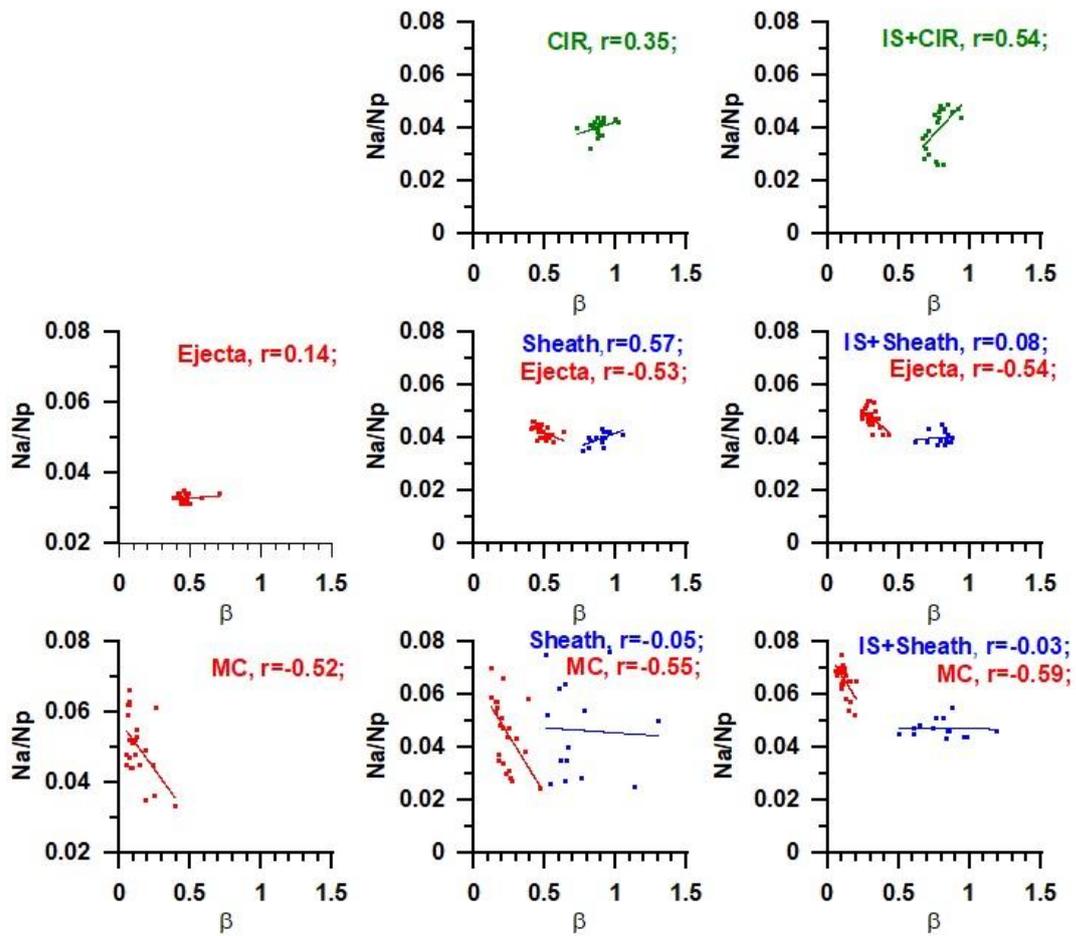

**Fig. 3.** *Dependence of helium abundance Na/Np on proton β-parameter in different SW types (see color legends in upper parts of each panel) for the 41-year set with helium measurements on an ACE satellite: points are the data obtained by the double method of superposed epoch analysis and presented in Fig.2; lines are their linear approximations; r is the correlation coefficient.*

Figures 4 and 5 present the similar data as Figures 2 and 3, but in contrast with Figures 2 and 3, they show results obtained for the 42-year set without ACE data. The results obtained for the two different data sets are very similar to each other.

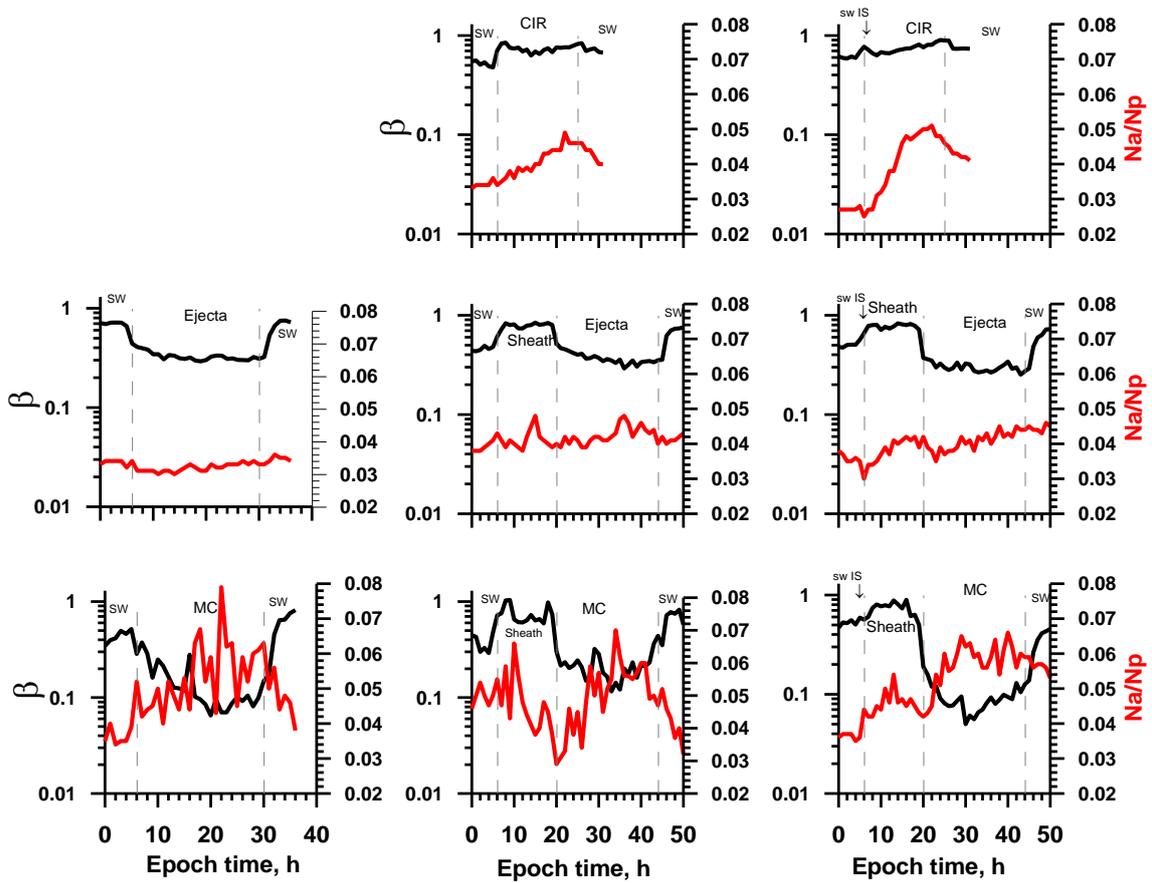

**Fig. 4.** *The same as Figure 2 for* the 42-year set without ACE data

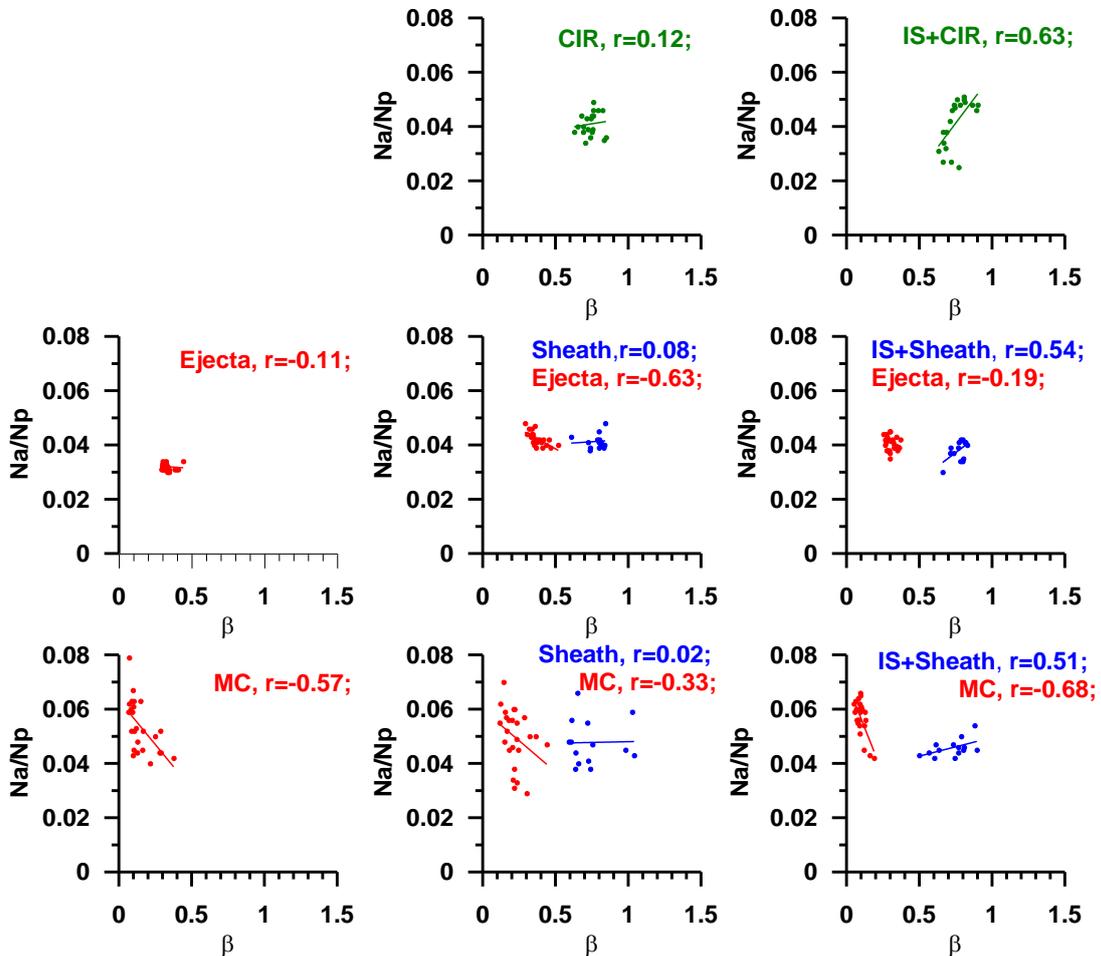

**Fig. 5.** *The same as Figure 3 for* the 42-year set without ACE data

## 4. Discussion and Conclusions

In this Section we discuss in detail the temporal profiles of helium abundance $N\alpha/Np$ and $N\alpha/Np$ vs. $\beta$-parameter dependences for eight usual sequences of SW phenomena presented in previous Section. In the beginning of this Section we would like to make two general comments.

First of all, it is necessary indicate that we use two different data sets with ACE measurements (Figures 2 and 3) and without ACE measurements (Figures 4 and 5), and comparison of results for two data sets shows that the exclusion of ACE data from consideration did not lead to any noticeable changes. This experimental result can be explained by suggestion that "a very poor correlation between the Wind/SWE/NLF Alpha/Proton density ratios and ACE/SWEPAM" is observed only at the scales of local-induced phenomena (see Introduction), and ACE data can be used for study at the scales of solar-induced phenomena.

Second, it should be noted that our results agree with the early obtained measurements: (1) the helium abundance in the steady slow SW streams from coronal streamers is lower than in the steady fast streams from coronal holes, and (2) the highest values of $N\alpha/Np$ observed in the magnetic clouds. Results for the disturbed SW types (CIRs, Sheaths and ICMEs) are discussed below.

### 4.1. Analysis of CIR data

<u>For CIRs without interplanetary shocks (ISs).</u> The helium abundance $N\alpha/Np$ varies approximately linearly from ~ 0.03 in the slow SW streams to ~ 0.04 in the fast streams. This is consistent with a hypothesis that the helium abundance in CIR corresponds to an intermediate value between the abundances in high-speed streams from coronal hole and slow streams from coronal streamer belt areas.

<u>For CIRs with ISs.</u> The helium abundance $N\alpha/Np$ in slow SW streams before CIRs is slightly lower than one for CIRs without ISs. It may be explained by the fact that velocity in slow streams before CIRs with ISs is lower than in streams before CIRs without ISs. In contrast with CIRs without ISs, the abundance $N\alpha/Np$ significantly changes and increases from ~0.025 up to ~0.05 throughout the interval.

Temporal profiles of $N\alpha/Np$ are sufficiently smooth in CIRs. Our results do not show any indications in favor of the hypothesis that several parameters abruptly change at so-called "interface" inside the CIRs (the boundary separating plasmas of fast and slow streams). Behavior of helium abundance at the interface requires additional selection of data and will be subject of further investigation.

### 4.2. Analysis of Sheath and Ejecta data

<u>For Ejecta events without Sheaths and without ISs.</u> The helium abundance $N\alpha/Np$ is equal to the abundance in the surrounding SW, almost does not vary within the interval and is ~ 0.03.

<u>For Ejecta events with Sheaths and without ISs.</u> The helium abundance $N\alpha/Np$ in Ejecta and in Sheaths is equal to the abundance in the surrounding SW, being almost constant within the intervals, and amounts to ~ 0.04.

<u>For Ejecta events with Sheaths and with ISs.</u> The helium abundance $N\alpha/Np$ in Sheaths is equal to the abundance in the SW before Sheaths, slightly vary within the intervals and is ~ 0.04. The helium abundance in Ejecta increases within the intervals from ~ 0.04 to ~ 0.05.

In both types of Sheaths before Ejecta (with and without ISs) the helium abundance is slightly higher than in CIR.

### 4.3. Analysis of Sheath and Magnetic Cloud data

The helium abundance $N\alpha/Np$ in MCs has large spread because of small number of measured phenomena being significantly higher than in Ejecta.

<u>For MC events without Sheaths and without ISs.</u> The helium abundance exceeds the abundance in the SW before MC (~ 0.03) and after MC (~0.047), strongly changes within the interval: there are an increasing trend (from 0.03 to 0.06) in the first part of interval, decreasing trend (from 0.06 to 0.05) in the second part, and increasing trend (from 0.05 to 0.055) in the short third part.

<u>For MC events with Sheaths and without ISs.</u> The helium abundance in Sheaths strongly changes within the interval (there is a decreasing trend) and varies from ~ 0.06 up to ~ 0.03 with the average value ~ 0.04. The helium abundance in MCs strongly changes within the interval (there are an increasing trend in the first part of interval and decreasing trend in the second part) and varies from ~ 0.03 up to ~ 0.07 with average value ~ 0.047.

<u>For MC events with Sheaths and with ISs.</u> The helium abundance in Sheaths slightly changes within the interval and is ~ 0.047. The helium abundance in MCs strongly changes within the interval (there are an increasing trend in the first part of interval and decreasing trend in the second part) and varies from ~ 0.03 up to ~ 0.075 with average value ~ 0.065.

It should note that the helium abundance $N\alpha/Np$ in MCs (3 bottom panels in figure 2) increases in the first half of interval and decreases in the second half, the β-parameter anti-correlates with $N\alpha/Np$. This result can be interpreted as the indication that spacecraft crosses a symmetrical structure in the solar wind which has maximal $N\alpha/Np$ and minimal β-parameter in the center of it.

### 4.4. Comparison of helium abundance $N\alpha/Np$ and *β*-parameter

The results of $N\alpha/Np$ vs. *β* relations slightly depend on the presence of shocks.

The increasing $N\alpha/Np$ vs. *β* trend is observed in the compression regions CIRs and Sheaths for cases with high magnitude of correlation coefficients, $|r| > 0.5$.

The decreasing trend is observed in both subtypes of ICMEs (Ejecta and MCs) for cases with high correlation coefficients, $|r| > 0.5$. The weaker dependence in Ejecta compared to MCs can be associated not only with the differences in the processes of their formation on the Sun, but also with the conditions of their observation near the Earth (spacecraft trajectory relative to ICME axe) (Yermolaev et al., 2017). This decreasing trend can be connected with solar conditions in the CME formation place: magnetic clouds (this type of ICMEs is the most clearly manifested in 1 AU) are formed in the areas (or altitude) of the solar atmosphere with high values of magnetic field and helium abundance. Since *β* is inversely proportional to magnitude of magnetic field, B, our results confirm results of Wang (2008, 2016) only for magnetic clouds, if a suggestion is correct that the magnetic field in SW phenomena correlates with magnetic field in their coronal sources. The obtained experimental result that MCs have maximal $N\alpha/Np$ and minimal *β*-parameter (maximal IMF magnitude, B) in the center of them support theoretical and experimental findings that heavy ions can play important role in the development of current structures in various space plasmas (Kistler et al., 2005; Zelenyi et al.,2006; Grigorenko et al., 2017). Figure 6 shows a schematic view of spatial axisymmetric distributions of helium abundance and *β*-parameter in the center of magnetic cloud. This property can be used for modeling CME/ICME structures.

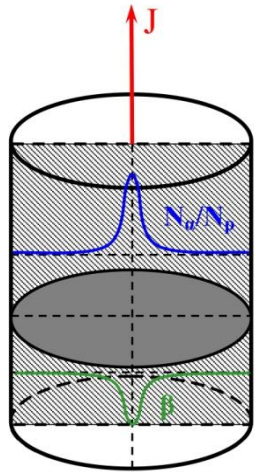

**Fig.6.** *Cartoon of cross section of MC: J (red arrow) - ion current, blue and green lines - Nα/Np and β distributions in the center cross-section of MC*

It should be emphasized that these trends are clearly observed only for the average profiles of *Nα / Np*, and in individual events have a wide spread. Therefore, *Nα / Np* data can be used to identify SW types only as a parameter that complements the main criteria, such as, for example, the *β*-parameter.

**Acknowledgements** We thank the OMNI database team for the opportunity to use data obtained from GSFC/SPDF OMNIWeb (http://omniweb.gsfc.nasa.gov). YY is grateful to the SCOSTEP "Variability of the Sun and Its Terrestrial Impact" (VarSITI) program for support of his participation in second General Symposium (VarSITI2017) in Irkutsk, Russia, 10-15 July, 2017. This work was supported by the Russian Foundation of Basic Research, project 16-02-00125.